# CONTAINERS AS THE QUANTUM LEAP IN SOFTWARE DEVELOPMENT

# Containers as the Quantum Leap in Software Development

The QLeap Team

**Abstract**. The goal of the project QLEAP (2022-24), funded by Business Finland and participating organizations, was to study using containers as elements of architecture design. Such systems include containerized AI systems, using containers in a hybrid setup (public/hybrid/private clouds), and related security concerns. The consortium consists of four companies that represent different concerns over using containers (Bittium, M-Files, Solita/ADE Insights, Vaadin) and one research organization (University of Jyväskylä). In addition, it has received support from two Veturi companies – Nokia and Tietoevry – who have also participated in steering the project. Moreover, the SW4E ecosystem has participated in the project. This document gathers the key lessons learned from the project.

**Contributors**. Iftikhar Ahmad, Teemu Autto, Teerath Das, Joonas Hämäläinen, Pasi Jalonen, Viljami Järvinen, Harri Kallio, Tomi Kankainen, Taija Kolehmainen, Pertti Kontio, Pyry Kotilainen, Matti Kurittu, Tommi Mikkonen, Rahul Mohanani, Niko Mäkitalo, Jari Partanen, Roope Pajasmaa, Jarkko Pellikka, Manu Setälä, Jari Siukonen, Anssi Sorvisto, Maha Sroor, Teppo Suominen, Salla Timonen, Muhammad Waseem, Yuriy Yevstihnyeyev, Verneri Åberg, Leif Åstrand.

# 1. Introduction

Continuous deployment has revolutionized the software development process by fundamentally changing how systems are designed, built, and released. Traditionally, software integration occurred on developers' workstations, following a structured, sequential process where each component was added in a carefully pre-planned order. However, with continuous deployment, integration happens dynamically, in real-time. This shift is particularly evident with the use of microservices, which serve as the modular building blocks of modern applications. In this new model, individual microservices can be integrated "on the fly" into a larger system, allowing developers to deploy updates and add new features quickly and continuously. This approach reduces the time between development and deployment, ensuring that end users receive regular, incremental updates rather than large, infrequent releases.

The shift to continuous deployment and microservices has significant implications for software architecture and product management. In traditional models, architecture and product planning could afford to be more rigid as changes happened at set intervals. Now, with continuous integration and deployment, both software architects and product managers must adopt new principles and strategies to keep up with the fast-paced nature of modern software. They must be able to handle constant changes without compromising the overall integrity of the software system. This requires a deeper understanding of microservice design patterns, as well as agile methodologies that accommodate rapid development cycles and incremental improvements. Moreover, this fast-paced environment demands enhanced communication and collaboration among teams, as each microservice's development impacts the system as a whole.

To support the continuous nature of modern software systems, new tools and technologies have emerged to facilitate seamless integration and deployment. Containerization, for instance, has become a critical element in the deployment process, with tools like Docker and Service Fabric providing a standardized environment for developing, testing, and deploying applications. Containers allow developers to package an application with all its dependencies into a portable unit that can run consistently across various environments. This not only enhances the reliability of deployments but also makes scaling easier, as containers can be added or removed based on demand. Alongside containers, orchestration tools such as Kubernetes have become essential for managing these deployments. Kubernetes automates the deployment, scaling, and operation of application containers, making it easier for teams to manage complex, distributed applications without excessive manual intervention.

In summary, continuous deployment has been a transformative shift in how software systems are built and maintained. The movement away from monolithic systems to microservice-based architectures has not only increased development speed but also introduced new complexities in terms of management and orchestration. As a result, software development teams must be well-versed in modern tools and practices to keep up with this rapidly evolving landscape. By leveraging container systems and orchestrators, developers and product managers can achieve the flexibility and scalability needed to meet today's demanding software requirements. Continuous deployment, supported by technologies like Docker and Kubernetes, enables organizations to deliver high-quality software at a rapid pace, ensuring that they remain competitive in an increasingly fast-paced digital world.

While the above techniques have been mainstream in cloud-based systems for some time, their role in systems that are not such has been less common. This has been the research gap that has been addressed by project Qleap, with its results presented in this document.

## 2. Background and Motivation

Over the past decade, continuous software development has become commonplace in the field of software engineering. New toolchains have emerged to manage the complexity in continuous deployment activity. Containers are a lightweight solution that developers can use to deploy and manage applications[1], often seen as a more lightweight alternative to Virtual Machines (VMs)[2]. Virtual Machines include the operating system where containers don't, allowing the containers to provide system resource usage advantages when compared against VMs[3].

The usefulness of containers is not limited to being a more lightweight version of Virtual Machines. One interesting feature of the containers is that they provide portability and, thus, modularity, making them suitable for working as software components[4] or as autonomous microservices[5]. When software systems grow, they encounter three problems:

1. Maintaining the software becomes harder.
2. Adding new features to the system slows down.
3. The resource requirements for the software grow.

One option to address these problems is to make systems modular[6]. In modular systems, software is split into smaller modules, and the full software systems are built by combining different modules[7]. Component-based software architecture and microservice architecture allow developers to build more modular software by plugging components together[8]. In component-based architecture, systems are created by connecting different software components[9]. Components are required when the system is compiled, and they are loaded when the system starts. Because of this, component-based systems do not help with the growing resource requirements but make maintaining the software easier.

---

[1] Paraiso, F., Challita, S., Al-Dhuraibi, Y., & Merle, P. (2016, June). Model-driven management of docker containers. In *2016 IEEE 9th International Conference on cloud Computing (CLOUD)* (pp. 718-725). IEEE.

[2] Dua, R., Raja, A. R., & Kakadia, D. (2014, March). Virtualization vs containerization to support paas. In *2014 IEEE International Conference on Cloud Engineering* (pp. 610-614). IEEE.

[3] Hoenisch, P., Weber, I., Schulte, S., Zhu, L., & Fekete, A. (2015, November). Four-fold auto-scaling on a contemporary deployment platform using docker containers. In *International Conference on Service-Oriented Computing* (pp. 316-323). Springer, Berlin, Heidelberg.

[4] Lau, K. K., & Wang, Z. (2007). Software component models. *IEEE Transactions on software engineering*, *33*(10), 709-724.

[5] Jaramillo, D., Nguyen, D. V., & Smart, R. (2016, March). Leveraging microservices architecture by using Docker technology. In *SoutheastCon 2016* (pp. 1-5). IEEE.

[6] Woodfield, S. N., Dunsmore, H. E., & Shen, V. Y. (1981, March). The effect of modularization and comments on program comprehension. In *Proceedings of the 5th international conference on Software engineering* (pp. 215-223).

[7] Card, D. N., Page, G. T., & McGarry, F. E. (1985). Criteria for software modularization. *Collected Software Engineering Papers, Volume 3*.

Similar to components, microservices are autonomous services that together fulfill a business requirement. Also, like component-based architecture, each microservice is required for the system to be fully functional. Since containers are not compiled as part of the software system, they could be used as a way to build plug-in-based architecture where containers-based plugins could provide new functionality into existing software, and they could be added and removed runtime[10]. Based on our observation, containers are used to build both component-based architectures and microservice architectures. Still, containers are often viewed as a way to lower resource requirements compared to Virtual Machines.

A recent systematic mapping study[11] reveals that while containers are commonplace, there are very different ways to use containers in practice. In addition, the study pinpoints the following observations:

- Containers are most often discussed in relation to cloud computing, performance, and DevOps. More than 50% of the papers discussed containers in the context of cloud computing, and performance-related aspects and DevOps were addressed in 45% of the papers.
- Containers are commonly used to modularize software systems, either through component-based architecture (28% of the papers) or through microservices architecture (26% of the papers).
- Docker is the dominant container design.

The study also reveals that there are multiple gaps or less-researched categories in using containers:

- Container security issues were not addressed.
- Legacy applications and their refactoring to container-based ones were not addressed.

---

[8] Voelter, M. (1999, July). Pluggable Component: A Pattern for Interactive System Configuration. In *EuroPLoP* (pp. 291-304).
[9] Crnkovic, I. (2001). Component-based software engineering—new challenges in software development. *Software Focus*, *2*(4), 127-133
[10] Birsan, D. (2005). On Plug-ins and Extensible Architectures: Extensible application architectures such as Eclipse offer many advantages, but one must be careful to avoid "plug-in hell.". *Queue*, *3*(2), 40-46.
[11] Koskinen, M., Mikkonen, T., & Abrahamsson, P. (2019, November). Containers in software development: a systematic mapping study. In *International Conference on Product-Focused Software Process Improvement* (pp. 176-191). Springer, Cham.

- While component-based and microservice architectures were common topics, no papers used containers as plugins or otherwise elements that would truly add flexibility to the systems.

In addition to technical changes in software development, containers will change the management of software products and create new business avenues for software vendors. From the business and management point of view, containers have profound effects. First, they add flexibility to software structures and their evolution over time, which requires new approaches to software product management. Second, they add flexibility to deployment and operations and also to collaboration between different stakeholders over the software development lifecycle. Third, they create a combinatorial explosion to product configuration alternatives. This makes software product management more challenging. Fourth, they provide new opportunities for developing software for distributed and shared platforms and ecosystems. Finally, there is little research on the use of containers outside the context of the web, and aspects such as regulatory compliance or AI explainability related to what happens inside containers is typically overlooked.

All these effects described above require new approaches and practices for managing software products and business. Unfortunately, there is practically no academic research work available for understanding and implementing these business and product management changes caused by containers. In our literature searches, we did not find any studies that investigated these challenges or provided solutions to them.

As a summary, Figure 2 presents a brief NABC analysis of the containers, microservices, orchestrators and the likes, and using and researching them in today's industry and research environment.

| Need | Benefits |
|---|---|
| • Continuous SE needs a flexible unit of deployment<br>• Deployed units need coordination<br>• Design space for deployment and coordination largely unexplored in domains other than web/cloud software<br>• Product management needs new approaches | • Well-considered use of containers simplifies development and deployment<br>• Flexible use of containers enables releasing features at faster speed than otherwise is possible<br>• Business opportunities from combinations and well-defined loosely coupled services<br>• Option to lead GAIA-X work |
| **Approach** | **Competition** |
| • Build a series of research sprints where each sprint addresses industry use cases in a collaborative fashion<br>• Seek business trials where companies share and/or subcontract their containers<br>• Foster industry-academia collaboration by enabling hybrid research teams | • International companies have implemented their own systems, based on their own standards and needs<br>• Presently, little effort to do the same for companies in collaboration collaborative way; GAIA-X is at present shaping up but monolithic platforms are getting a head start |

Figure 2. Project content NABC diagram.

## 3. Research Themes and Key Results

### 3.1 Literature Study and Multivocal Study on Container Orchestration

In the early 2010s, many new ideas for the software development paradigm, such as agile development, were introduced. The software technology was still not mature enough to offer resources for abstracting fine-grained software components. Then came a lightweight virtualization technology known as the containerization technology that was made popular mainly by the Docker software. It provided an encapsulation mechanism for bundling applications, their runtime, and all their dependencies, making it possible to run applications securely and on any platform that can handle containers.

Based on the level of abstraction, two major virtualization techniques can be defined: virtual machine (VM) -based virtualization, and container-based virtualization. The VM-based virtualization virtualizes the entire operating system (OS), and it needs an underlying hypervisor to function. The container-based virtualization uses techniques, such as namespaces and cgroups, within the OS kernel to isolate access paths for the resources, and it does not need the hypervisor to function. The Docker software made it straightforward to seamlessly move applications among heterogeneous (different from each other) environments.

There was a need for an orchestrator study because the containerization technology has significantly evolved during the late 2010s and early 2020s, but not much research has been done on the orchestrators. The widespread use of containers in core, edge, and far-edge computing within distributed cloud environments introduces numerous challenges. Addressing these challenges requires advanced orchestration solutions that can efficiently manage the deployment, scaling, monitoring, and security of containers. However, achieving this level of orchestration is not straightforward, as it introduces various types of challenges across different layers of the system. Rather little is still known about the state-of-art in the container orchestration field, *suggesting a research gap*.

To address this identified research gap, the objective of this research was to systematically identify and categorize the challenges, solutions, strategies, and architectural designs of container orchestration in distributed cloud systems. We aimed to explore various aspects of container orchestration, including strategies for core, edge, and far-edge computing, and examine their quality attributes such as scalability, performance, security, and fault tolerance. By employing a systematic mapping study approach, we provide a comprehensive analysis of 86 selected studies, contributing to the development of more effective orchestration solutions that can address the demands of distributed cloud environments.

Petersen et al. (2008)[12] described the process of systematic mapping study as follows:

---

[12] Petersen, Kai, et al. (2008). Systematic Mapping Studies in Software Engineering. *12th international conference on evaluation and assessment in software engineering (EASE)*.

      a. Definition of Research Questions (Research Scope)
      b. Conduct a Search for Primary Studies (All Papers)
      c. Screening of Papers for Inclusion and Exclusion (Relevant Papers)
      d. Keywording of Abstracts (Classification Scheme)
      e. Data Extraction and Mapping of Studies (Systematic Map)

*"Collective Intelligence for the Internet of Things"*[13] is one of the journals identified for publishing the results of the multivocal study on container orchestration. We plan to submit the results in the coming months (possibly during early 2025).

**Results.** The literature study[14] had 52 papers reviewed, and the most common keywords were extracted for the multivocal study. The following list contains the major trends and active areas of research in container orchestration found by the preliminary literature study:

- Service Discovery and Networking
- Resource Allocation
- Load Balancing
- Scheduling and Scaling
- Zero-Downtime Updates and Rolling Deployments
- Health Checks and Self-Healing
- Interoperability and Integration
- Replication
- Monitoring and Logging
- Security Features

*The literature study confirmed the research gap* we were expecting to see. Several research questions were formed for the multivocal study to map a wider field of study. After forming the five research questions, the following seven academic databases were queried for the wider article collection: IEEE Xplore, ACM Digital Library, Scopus, Web of Science, SpringerLink, ScienceDirect (Elsevier), and Wiley Online Library. A generic search string *"container orchestration" AND "distributed cloud" AND challenges AND solutions AND "quality attributes" AND "architecture design"* was tailored for each of the databases with the aid of ChatGPT.

**Conclusion.** While container technologies have matured due to the move towards more platform-independent and resource-efficient application deployment solutions, effective orchestration has not yet received sufficient attention, especially in distributed cloud systems across core, edge and far-edge environments. Due to the multitude of problems and solutions offered for the listed areas of application, it may be hard to find suitable candidates for containerization. This study aimed to fill in the research gap, offering new information about the challenges, solutions, strategies, and architectural designs required to manage containerized applications at scale.

---

[13] Call for papers - Internet of Things | ScienceDirect.com by Elsevier: *Collective Intelligence for the Internet of Things*.

[14] Järvinen, V., Sorvisto, A., Heinonen, H., Paavonen, A.-S., Karthikeyan, D. K., Oliver, I., Waseem, M., Mäkitalo, N. & Mikkonen, T. (2024). Container Orchestration Taxonomy. Poster presented at: *QLeap: Sprint 8 Review & Final Sprint Planning.* (2024, August, 27). Jyväskylä, Finland.

Our objective was to systematically identify key challenges in container orchestration and explore solutions that address deployment, scaling, monitoring, and security demands within distributed cloud environments. We conducted a systematic mapping of relevant studies, evaluating core quality attributes such as scalability, performance, security, and fault tolerance to build a detailed understanding of current orchestration practices. Our preliminary analysis of the research revealed primary research themes, including service discovery, resource allocation, load balancing, scaling, rolling updates, health checks, interoperability, replication, monitoring, and security features. The results of the wider multivocal study provide a robust framework that will support the development of more efficient and resilient orchestration solutions in distributed cloud environments. The state-of-art taxonomy of container orchestration will be submitted for publication in early 2025.

## 3.2 Containers in Multi-Cloud Context

In this use case, we investigated containerization in a multi-cloud environment, identifying the roles, strategies, and challenges associated with containerization in multi-cloud settings[15]. Our findings propose and develop theoretical frameworks to address challenges related to automation, monitoring, deployment, and security by offering structured solutions. These frameworks are designed to enhance the efficiency of containerized applications in multi-cloud environments, providing clear guidance on managing resources, ensuring scalability, securing workloads in dynamic, distributed cloud settings, and addressing various other issues. Additionally, we identified 74 patterns and strategies (successful practices) across four themes—Scalability and High Availability, Performance and Optimization, Security and Privacy, and Multi-Cloud Container Monitoring and Adaptation. Furthermore, 47 tactics addressing 10 quality attributes were identified, providing a solid foundation for practitioners to improve application deployment, security, and automation processes. By adopting these strategies and frameworks, practitioners can tackle critical challenges and enhance the efficiency and reliability of containerized applications in multi-cloud environments, resulting in better workload management, improved security, and reduced operational complexities.

The outcomes of this use case are currently under review in ACM Transactions on Software Engineering and Methodology (TOSEM), and a preprint has been published online[16].

## 3.3 Large Language Models in Software Engineering

As the importance of large language models (LLMs) for software engineering became apparent during the project, some survey work on LLM usage was started. The first survey produced was on running LLMs locally and focused on quantization techniques to reduce the model memory footprint and improve execution speed. Resource-efficient fine-tuning of models for company-specific use cases was also investigated.

---

[15] Waseem, M., Ahmad, A., Liang, P., Akbar, M. A., Khan, A. A., Ahmad, I., Setälä, M.. & Mikkonen, T. (2024). Containerization in Multi-Cloud Environment: Roles, Strategies, Challenges, and Solutions for Effective Implementation. *arXiv preprint arXiv:2403.12980*.

[16] Waseem, M., Ahmad, A., Liang, P., Akbar, M. A., Khan, A. A., Ahmad, I., ... & Mikkonen, T. (2024). Containerization in Multi-Cloud Environment: Roles, Strategies, Challenges, and Solutions for Effective Implementation. *arXiv preprint arXiv:2403.12980*.

The second survey expanded on LLM use for company-specific use cases by investigating retrieval augmented generation (RAG). Different RAG methods and adjacent technologies like vector databases were surveyed. In parallel a test case of using Azure cloud platform for fine-tuning foundation models using techniques investigated in the first survey.

The third and last survey was on the current state of software generation with LLMs. The topics covered were techniques to ensure code correctness, class-scale code generation and repository-scale generation. The code-correctness approaches found largely relied on coupling an LLM with a formal verification tool. The class-scale generation survey found that LLMs had trouble generating these larger blocks of code compared to single functions. Repository scale generation refers to code generation tasks too large to fit in the LLM context window. The survey found two main approaches to this problem:

- The first approach was generating a skeleton or plan for the whole code and then generating the code piece by piece.
- The second approach was using an agent-based system, where the generation task is split into smaller pieces and each piece is assigned to an LLM agent.

The findings were presented during sprint meetings. In the first survey, a technique called QLoRa was found as the most popular for reducing fine-tuning requirements. This technique was then tested when exploring Microsoft Azure cloud platform for fine-tuning large language models. It enabled fine-tuning of large models on lesser resources and despite some issues, the platform was found to be relatively easy to use. In the final survey it was found that current LLMs while good at function-level coding tasks, struggle with class-level tasks. It was also found that repository scale generation is a big challenge due to the limited input length of LLMs. Two research approaches were identified: first was top-down generation, where you first generate a skeleton of the code and then generate smaller pieces to fill it in. The second was using an agent-based system where you have some central logic deciding the smaller tasks and running multiple agent LLMs to generate the subproblems.

## 3.4 Ecosystem Formation, Governance, and Containerization

This study explored the formation and governance of software ecosystems, focusing on their role in enabling collaboration, resource sharing, and technological innovation, such as utilizing container software. Our findings highlight how ecosystems reduce development risks, facilitate knowledge exchange, and support rapid responses to environmental demands, such as competition, changing user preferences, or other critical conditions[17]. Specifically, software ecosystems centered around platforms or related technologies streamline resource integration and adaptation, creating synergies among participants. According to our discoveries during the QLeap, containerization enables the building of complex, distributed systems that, in turn, demand implementing systematic and structured ecosystem governance.

---

[17] The manuscript titled "Unifying a Public Software Ecosystem: How Omaolo Responded to the COVID-19 Challenge". *arXiv preprint* https://arxiv.org/abs/2410.00668

**Ecosystem governance** involves establishing and managing shared processes, rules, and models for effective collaboration. Unlike rigid contract-based networks, ecosystems are usually dynamic and interdependent in their nature, driven by a shared vision. Changes, such as API updates or introducing a new software, can trigger ripple effects throughout the system in a long period of time, underscoring the importance of robust coordination.
In the QLeap project, we studied two different ecosystems and their governance structures:

- An open-source platform ecosystem coordinated by Vaadin, who was in the process of integrating containers, and
- An innovation ecosystem SW4E focused on software innovation, such as leveraging container technology, facilitated by DIMECC.

In collaboration with our two project partners, we collected data on ecosystem structures, governance mechanisms, and participant dynamics. This collaboration has resulted in ongoing work on two manuscripts. The first manuscript[18], examines the challenges and strategies for maintaining viability and sustainability in the Vaadin open-source platform ecosystem, using it as a practical case for implementing a domain-specific modeling language for digital ecosystem governance. The second manuscript[19], explores how modeling complex ecosystem dynamics can guide ecosystem formation and support value co-creation within the SW4E innovation ecosystem.

**As initial findings** on these two use cases, we identified governance best practices essential for maintaining sustainability, fostering community building, promoting knowledge exchange, and supporting data-driven decision-making in software ecosystems. Key practical insights include:

- Designing and continuously developing a governance framework requires structured and systematic approaches for a clear, yet detailed overview of the situation. This involves analyzing governance perspectives, examining relationships between ecosystem elements, and effectively communicating these insights to stakeholders.
- There is a need for shared guidelines and jointly agreed practices to facilitate resource sharing, buy-versus-build decisions and access to shared knowledge. This, together with trust in the technological infrastructure of ecosystems, enables co-creation of value, for example in innovation activities, among ecosystem participants.
- Governance frameworks need to take into account regulatory and legislative considerations when seeking to improve competitive advantage with emerging technologies such as containers. Coordinating and automating compliance and sharing information about relevant regulatory frameworks in the ecosystem were examples of concerns raised during the project.

---

[18] The publication's working title is "Implementing a Domain-Specific Modeling Language for Designing Collaborative Ecosystems and Their Governance."
[19] The publication's working title is "Use Case Study for Modeling Business Value in an Innovation Ecosystem."

In the QLeap project, we found that container technology has the potential to significantly enhance ecosystem dynamics by enabling seamless system interactions, facilitating resource reuse, and lowering participation barriers. Containers provide a robust compatibility layer that ensures platform independence, allowing more participants to access complementary resources and generate value within the ecosystem. This modularity supports horizontal development and strengthens value chains by improving scalability, flexibility, and developer productivity. Additionally, containers enhance security and ecosystem reliability by offering natural isolation boundaries and ensuring consistent operations across platforms through standardized, portable environments. Further research is needed to explore container technology's role in fostering ecosystem value co-creation and governance frameworks, especially including decision-making, for distributed systems based on containers.

## 3.5 Containers and Security Issues

Container risks and vulnerabilities are critical aspects of container security. Risks and vulnerabilities have significant implications in the performance and the availability of the containerized services. A mapping study was conducted to summarize the current state of the art on container risks and vulnerabilities. The findings from the mapping study point to configuration flaws as the main cause for risks and vulnerabilities in container systems. Additionally, it highlights best practices in container security, offers mitigation strategies to prevent implementation errors and attacks, and presents a range of tools to sustain container system security. The findings also found that the causes for risks and vulnerabilities stem from misconfiguration issues or configuration flaws. The mapping study manuscript is currently under review in IST Information and Software Technology, but a preprint is available[20].

As found from the Mapping study, researchers made efforts to investigate container security theoretically and experimentally. Meanwhile, software practitioners have also developed practices to improve and maintain container security based on their work experience. Semistructured interviews were conducted with practitioners across various domains. The interviews aimed to explore various opinions on container security issues, causes, implications, tools, and practices used in real containerized projects. The findings reveal repeating patterns among various domains using containers for deploying software. Modelling the patterns and their relationships reveal the strengths and weaknesses of container security in practice. Strengths include a comprehensive understanding of security issues, reliance on tools and automation, awareness of security dependencies, and consideration of non-technical factors. Conversely, weaknesses encompass the lack of systematic knowledge, guidelines, and standards, uncertainties regarding practice improvements, and resilience time. The study already accepted and presented in Software Engineering and Advanced Applications (SEAA2024)[21].

---

[20] Sroor, M., Das, T., Mohanani, R., & Mikkonen, T. (2024, February). A Systematic Mapping Study on Software Containers Risks and Vulnerabilities. https://dx.doi.org/10.2139/ssrn.4741002

[21] Sroor, M., Mohanani, R., Das, T., Mikkonen, T., & Dasanayake, S. (2024, June). Practitioners' Perceptions of Security Issues in Software Containers: A Qualitative Study. In *Software Engineering and Advanced Applications* (SEAA2024). (https://www.researchgate.net/profile/Rahul-Mohanani/publication/381780830_Practitioners'_Perceptions_of_Security_Issues_in_Software_Container

In addition to the two previous papers, a survey was conducted to examine company practices and typical testing tools. The findings provide an overview of the current testing approaches employed by companies, the significance of tailored testing practices, and the consensus on multi-phase testing processes. It also illustrates the importance of testing, with the rise of security and vulnerability issues and challenges. The data collected was used for a master's thesis entitled "Detecting Anomalies by Container Testing: A Survey of Company Practices and Typical Tools"[22]. The survey data was further analyzed and resulted in a research paper published in Profes 2023[23]. The paper provided managerial and practitioner recommendations to improve container system testing approaches.

Currently, we have two papers ongoing. The first paper will conduct a survey of expert opinions and ideas regarding the prioritization of container vulnerabilities within container environments. The primary objective is to define a set of minimum security requirements for container systems, with a focus on mitigating high-risk vulnerabilities. The second paper is in its early planning stages and will explore the application of current security best practices to container systems.

## 3.6 Zero Downtime Techniques

In this use case, we explored various aspects of achieving zero downtime in cloud applications from the perspective of software practitioners. We conducted a semi-structured interview study with 16 developers to analyze key zero downtime strategies, including tools, techniques, architectural patterns, and the challenges involved. The study also examined best practices and requirements for both new cloud applications and those not initially designed for zero downtime. Our findings identified 12 distinct tools, techniques, and architectural patterns commonly used by developers to ensure zero downtime. The most widely adopted method is rolling updates (also known as gradual updates), followed closely by the blue-green deployment technique. Additionally, practitioners highlighted the importance of microservices architecture in achieving zero downtime. Other popular techniques include canary deployment and the use of feature flags, with Azure being the most commonly used tool among practitioners for achieving zero downtime. We compiled a catalog of 21 unique challenges practitioners face when implementing zero downtime. The most common and recurring issue was database incompatibility, followed by concerns around security exposure. Data and database incompatibility issues arise when changes made to the application during local development require corresponding updates to the database, creating a dependency between them. If the new version of the database is deployed while the old version of the application is still running, this mismatch can lead to errors. Additional

---

s_A_Qualitative_Study/links/667eba202aa57f3b825cff6a/Practitioners-Perceptions-of-Security-Issues-in-Software-Containers-A-Qualitative-Study.pdf)

[22] Timonen, S. Detecting Anomalies by Container Testing: A Survey of Company Practices and Typical Tools. (2023, May). Master's Thesis in Information Technology, University of Jyväskylä Faculty of Information Technology. http://urn.fi/URN:NBN:fi:jyu-202305303334

[23] Timonen, S., Sroor, M., Mohanani, R., & Mikkonen, T. (2023, December). Anomaly Detection Through Container Testing: A Survey of Company Practices. In *Product-Focused Software Process Improvement* (PROFES 2023). Lecture Notes in Computer Science, vol 14483. Springer, Cham. https://doi.org/10.1007/978-3-031-49266-2_25

challenges include transitioning legacy systems to zero downtime, a lack of developer skills or knowledge in this area, and budget or cost constraints, among others.

The findings from this use case are documented in a research paper[24], which has been submitted for review to the International Conference on Software Engineering - Software Engineering in Practice 2025 (ICSE - SEIP).

### 3.7 Building of the Practical Demo

To make things practical, a demonstration setup was built at the University of Jyväskylä to test a compact High Availability (HA) cluster with robust security and efficient container orchestration. Challenges included implementing containerized software and deprioritizing features like SIEM and secure enclaves due to resource and complexity constraints. The setup consisted of a four-node distributed cluster with three HA nodes and a hot-swappable spare.

Dell OptiPlex Micro machines (14th gen i5, 32GB DDR5) were chosen for portability, efficiency, and vPro-enabled lights-out management. While separate management networks were recommended, hardware limitations required using VXLAN for secure virtual networking. Machines ran Alma Linux, with tweaks for container workloads and Salt/Ansible for configuration management. K3s was selected as the orchestration system due to its lightweight Kubernetes-based design. Security features included Shamir's Shared Secret for data at rest, WireGuard for data in transit, and secure boot with MOKs. Post-Quantum Cryptography (PQC) and measured boot were evaluated but not fully implemented. High encryption demands were met using TPM-based hardware RNGs, supplemented by custom random-seeding methods for resource-limited devices.

Monitoring shifted from Netdata to Grafana/Prometheus due to licensing changes, with additional custom metrics for anomaly detection. Certificate management used cert-manager with step-ca and a robust offline backup strategy. Lightweight SIEM via CrowdSec was implemented, alongside Kata Containers for privileged process isolation and testing future Confidential Containers (CoCo) support.

The cluster followed GitOps practices, with Flux CD for deployments and GitLab integration for pipelines. Distributed block storage supported data redundancy. Build pipelines included signature verification and Trivy-based scanning for supply chain security, aligning with SLSA level 2 standards and preparing for level 3.

---

[24] Das, T., Kumar, R., & Mikkonen, T. (2024, October). Zero Downtime in Cloud Computing: A Qualitative Study from the Lens of Software Practitioners. International Conference in Software Engineering - Software Engineering in Practice (SEIP), 2025 (Submitted). https://www.researchgate.net/publication/385145830_Zero_Downtime_in_Cloud_Computing_A_Qualitative_Study_from_the_Lens_of_Software_Practitioners

## 4. Company Contributions

### 4.1 Bittium

Bittium ([www.bittium.com](www.bittium.com)) had the following targets for the project which were more or less achieved in full. *First objective* was to research the *aspects of containers in Embedded SW context and various domains* (generic/domain specific applications), combining security & traceability aspects and ML/AI Ops tooling towards RegOps. The approach is currently used more and more widely and described in the use case adopted by Bittium.

*Second main objective* was to develop the direction of "Running R&D in containers" to scale business assets and reduce potential technical debt. For this a very specific enabler was Value Stream visualization approach which was itself containerized but enabled transparency with the other tools adopted to potential technical debt and therefore seen as a very important aspect. The development of container architecture of scaled containers solutions was combined with used JIRA & JIRA structure tool environments in connection with the containers itself as well as for example X-Ray testing system.

Third main objective was the adaptation of the efficient container approach for the next gen products and solutions (practical example cases) with various on-prem and hybrid environments. This was specific for Bittium as some of the development environments must be currently on-prem due to requirements set by e.g. governmental customers.

The results derived from the first key target were the following. Traceability of the containers was being developed including all the data elements in it, and as an example from Bittium approach and architecture for over 200 containers was presented in QLeap. This will enable traceability of the containers and all the data elements in it as well as lifecycle of containers up to tens of years. With the approach the lifecycle of containers up to tens of years required by the use case was planned. The security aspects (connection of vulnerability management and anomaly detection, security scenario testing (MITRE, OWASP etc…) are connected to the deployment phase later on.Visual management (Real-time KPIs, technical debt) is in place some of the use cases, wide scale adoption is being studied for the future development.

For the second objective "*R&D in the containers*" Bittium demoed the impacts of the activities with Value Stream Management (VSM) approaches in the Sprint review of QLeap (Konttihyppy). The VSM tool is itself containerized and we have now also incorporated the VSM tool to other environments as the restricted one to be able to measure KPIs like Queues, Velocity, Load, Flow/Cycle time, Efficiency and Issue distribution. Other results include the following like automated container generation with DevOps Pipeline now in use. It was also noted that technical debt visualisation requires process change to make it visible in the tool chain. The VSM tool enables automatic metric generation, as easy as possible implementation of the tool from the container.

4.2 M-Files

During the project, M-Files has taken a holistic view on how it develops and provides cloud services to its customers by creating a 3-year plan outlining the company's strategy to enhance its cloud services from 2024 to 2026. The plan emphasizes the need for a scalable, enterprise-ready cloud infrastructure that can support a significantly larger customer base without proportional increases in operational costs.

Key initiatives include the development of an autonomous infrastructure, self-service capabilities, improved security measures, and a transition to a future-proof architecture based on Azure Kubernetes Service. The roadmap includes specific milestones for product enhancements, self-service capabilities for customers and partners, and operational efficiencies to ensure sustainable growth.

M-Files will focus on implementing a Platform Engineering approach to simplify technical complexities and improve service quality while reducing operational costs. The company plans to develop the M-Files Cloud Run platform to automate cloud-related processes and manage the entire lifecycle of applications and infrastructure. Future enhancements will include improved cloud architecture, high availability options, and monitoring capabilities. Additionally, the plan addresses the need for specialized expertise in areas such as Kubernetes and Azure governance to support enterprise-level service delivery.

Containerization and utilization of scalable, serverless cloud services are crucial to the implementation of this 3-year plan. When aiming at full automation of running and scaling the cloud services, this cannot be achieved without fully isolated, containerized services.

Several concrete outcomes that served the purpose of creating the 3-year plan were achieved during the project. Some of these outcomes were presented and demonstrated in the project meetings and such examples include:

- Cloud-optimized PDF conversion. M-Files planned and implemented a technical proof of concept for a cloud architecture that allows to move resource intensive workloads from the core M-Files system to automatically scalable Azure cloud services. This is crucial when processing large amounts of documents in a fast and cost-effective manner. The solution utilized technologies such as Azure Functions, Azure Message Bus and Azure Blob Storage.
- Zero-downtime enablers. In every zero-downtime initiative, it is vital that all centralized databases maintain their backward compatibility when the application containers are updated on the fly to new versions. M-Files planned and developed a technical proof of concept that does exactly that for the relational database.
- Azure Kubernetes Service (AKS). During the project timespan (but not as part of the project itself), M-Files successfully migrated all its cloud customers from a previous generation cloud platform to Azure Service Fabric. Even though this platform has worked well, it has still become evident that for the longer-term scalability for the enterprise-level cloud services, this platform needs to be further changed to AKS. During the project, M-Files conducted several technology trials to ensure suitability of AKS for this purpose. Trials were successful, and now the plan is to move ahead at full speed with the AKS migration during 2025.

To summarize, from M-Files' perspective QLeap has been an important and successful project that has significantly helped to speed up moving to the next level with M-Files Cloud. In addition, successful technology trials and collaboration with other project partners have reduced technological risks that are always present in fundamental and rapid changes such as these.

4.3 Nokia

From Nokia's side, this project has been in a key role to create new software capabilities based on the containers that help to clarify some essential approaches on edge computing and multiple and/or hybrid cloud environments. These have been a central role in Nokia's Veturi program and the developed new technical capabilities can be used e.g. on life cycle management of the wireless software-based systems in the different industrial domains, and they can help to select the most suitable architectural choices for the future edge solutions and applications especially in the multi-cloud environment.

Nokia has been part of the project steering activities and contributed for example result dissemination by providing opportunities for the project team to share a keynote presentation by Prof. Tommi Mikkonen as part of Nokia's Technology Tuesday webinar where 700+ Nokia participants were able to see the latest results of the project. In addition, project and Nokia teams have collaborated to jointly identify the industry challenges and potential solutions. Nokia also facilitated discussions with Nokia's senior R&D leaders in Spring 2024. Another event was arranged on Nov 26th 2024 at Nokia HQ campus to demonstrate demos for Nokia's technology experts and R&D engineers.

4.4 Solita/ADE Insights

This project has been very beneficial for Solita. Our work items provide us with a lot of new knowledge and we have already used the knowledge in our business projects. Also during this project Solita reorganized its corporate structure, and this project was transferred to a newly established entity ADE Insights Oy - currently results are shared between Solita Oy and ADE Insighs Oy.
Solita has 7 main own work items, but 2 of those were very small and they are not described in this document. Also "Containers in edge" were small but relevant, in that package we created a model on how to use MLOps with IoT Edge. The result of that work is documented in an article that was published & presented in ACM eSAAM '23. https://dl.acm.org/doi/10.1145/3624486.3624496.
More detailed information of work & results from the bigger work items.

Solita WI-1: Scale-out Containerized Software Development Process

Transitioning from monolithic environments to container-based architectures significantly improved scalability, agility, and developer productivity. Key achievements include:

- Improved Build Granularity: Programmatic generation of build configurations reduced manual work and errors, accelerating development cycles.
- Parallel Test Execution: Enabled by containerized infrastructure, reducing resource bottlenecks.
- Dynamic Caching Strategies: Optimized build and scan times in serverless environments, enhancing efficiency.
- Cross-Platform Development Support: Enhanced support for diverse hardware configurations (e.g., ARM) to ensure reliability across environments.

### Solita WI-2: Security

Securing containerized applications and environments was addressed by automating vulnerability management and consolidating container images. Key findings include:

- Image Standardization: Reduced image variations to streamline security updates and minimize maintenance overhead.
- Automated Vulnerability Management: Introduced metadata-driven scanning and impact analysis to improve response times and reduce manual auditing.
- Multi-Tool Security Scanning: Developed a scalable system supporting diverse scanning tools across multi-cloud ecosystems.
- Supply Chain Threat Mitigation: Conducted continuous security testing (e.g., penetration tests) to adapt to evolving threats, improving overall system security.

### Solita WI-3: Cost and Performance Optimization

Cost-saving measures and performance enhancements were achieved by optimizing resources and deployment processes. Key insights include:

- Cost Awareness: Teams gained insights into the cost implications of daily decisions, leading to better practices.
- Diverse Instance Utilization: Leveraged multi-CPU architectures (e.g., ARM) to reduce testing and deployment costs.
- Continuous Deployment Improvements: Enhanced CI/CD pipelines to support more iterations per day and minimize downtime during updates.
- Memory Optimization: Reduced memory requirements for containerized components, especially Java-based services, to decrease operational costs.

Solita WI-4: Multi-Tenant

Multi-tenant architectures were analyzed to optimize scalability and cost efficiency. Key results include:
- Migration to Multi-Tenant: Carefully tested migration from single-tenant to multi-tenant services, achieving reduced downtime during updates.
- User Behavior Analytics: Developed a prototype for collecting and analyzing service metadata across scaled-out environments, improving system understanding.
- Open Source Alternatives: Evaluated the trade-offs of replacing open source software with in-house developed microservices to improve performance and user experience while reducing maintenance complexity.

Overall, the project can be considered a success, and the research outcomes have been implemented as part of the ADE product through in-house development efforts.

## 4.5 Tietoevry Finland Oy

Digital services are becoming an essential part of everyday life. It is crucial that these services are available, reliable, and protected from various vulnerabilities and attacks. Additionally, digitalization brings an increasing amount of personal data that needs to be processed in different contexts, whether in single or multi-cloud environments. Therefore, it is vital for modern digital societies to protect digital products and services while also making various types of data shareable among multiple stakeholders. This requires fine-grained control over data, not only for processing but also for enriching it with metadata.

It is vitally important to create secure, reliable, and scalable digital products & services for data processing in a dynamic environment. For Tietoevry, containers play a fundamental role in secure, scalable data sharing with fine-grain control over different capabilities listed below.

1. Isolation and Consistency: Containers encapsulate an application along its dependencies, ensuring that it runs consistently across different environments.
2. Portability: Applications can be deployed on different cloud services and even on private clouds.
3. Scalability: Containers can be scaled up or down to handle varying loads. Orchestrators like Kubernetes manage the scaling process by adding or removing container instances based on demand.
4. Resource Efficiency: Containers share the host system's kernel and resources, making them more lightweight compared to virtual machines. This can enable more containers on the same/similar hardware.
5. Microservices Architecture: Container makes it easy to develop, deploy and maintain many services efficiently. This enables faster development, deployment and efficient maintenance of the digital services.
6. Continuous Integration and Continuous Deployment (CI/CD): Containers integrate well with CI/CD pipelines. This automation reduces the risk of human error and speeds up the release cycle.

7. Zero-Downtime Deployments: Containers inherently support rolling updates and blue-green deployments, which enable zero-downtime updates.
8. Health Checks and Self-Healing: Container orchestrators such as Kubernetes provide mechanisms for health checks and self-healing. Liveness and readiness probes monitor the health of containers, and if a container fails, the orchestrator can automatically restart it or replace it with a healthy instance.
9. Security and resilience: Containers enhance security by isolating applications from each other and the host system. Additionally, container images can be scanned for vulnerabilities before deployment, and runtime security tools can monitor container behavior.
10. DevOps and Collaboration: Containers facilitate collaboration between development and operations teams. Developers can package their applications with all dependencies, ensuring that they run consistently in any environment. Operations teams can manage and scale these containers efficiently, fostering a DevOps culture.

## 4.6 Vaadin

Vaadin has participated in the project to investigate ways of supporting customers building applications using a micro-frontend architecture. This kind of architecture uses the same principles as for microservices but applies them for the web-based UI layer of the application. Just like microservices for backend functionality, micro-frontends also rely on containerization – first with conventional server-side containers running microservices to provide data to the frontend and additionally also with container-like isolation capabilities in the web browser to avoid accidental interference between separately developed and deployed parts of the front end. Vaadin's main project goal was to develop a reference implementation for a Vaadin-based micro-frontend architecture and validate this implementation with pilot customers. This goal was not fully achieved due to circumstances outside the scope of the project.

First, Vaadin was, independently of the QLeap project, updating its architecture to use a frontend toolchain based on Vite ([https://vite.dev/](https://vite.dev/)) in favor of the previously used webpack ([https://webpack.js.org/](https://webpack.js.org/)) tool. While Vite provides a noticeable improvement in the overall developer experience, it turned out that its module federation support has so far been lacking. Module federation allows micro-frontend modules to dynamically load transitive dependencies from a shared common module so that each micro-frontend wouldn't have to load its own copy of those dependencies. This capability is essential for a micro-frontend architecture due to the way Vaadin's web components use Custom Elements which can only be registered once in each browser window. Instead of developing a reference architecture based on module federation, Vaadin ended up building several prototypes to explore the trade-offs of different alternatives to module federation.

The second external circumstance was the overall trend shift away from seeing microservices as a universal approach to application development. After the initial microservice hype, the ecosystem has started gaining a deeper understanding of how to balance between the benefits of team independence and the overhead of managing a complex distributed system. This perspective has led to the rise of a modular monolith architecture where modules are developed by independent teams while the application is still deployed as a single unit. Vaadin's current products provide a good fit for this kind of architecture which has reduced the necessity of developing a full-featured micro-frontend reference architecture based on module federation. At the same time, the shift in the ecosystem has reduced customer interest for micro-frontends so that no suitable pilot customer has been found for evaluating the developed prototype architectures.

Despite the complications with Vaadin's main project goal, there have still been multiple secondary benefits from participating in the project. Vaadin has investigated ways of making its main development framework function better in clustered environments and for various embedded UI use cases, added SBOM generation for its own product builds to be used as input for container security solutions, and prototyped products for container orchestration and application-level SBOM generation and analysis.

## 5. Discussion and Future Work

During the project time, software containers have become a cornerstone of modern application development and deployment, driven by their ability to ensure consistency, scalability, and efficiency. Tools like Docker and Kubernetes have gained widespread adoption, enabling developers to encapsulate applications and their dependencies in lightweight, portable units that run reliably across diverse environments. This surge in popularity is largely due to the acceleration of cloud-native development, the growing complexity of microservices architectures, and the need for rapid deployment in DevOps pipelines. Organizations, including in particular the ones that have participated in the project, have embraced containers to reduce overhead, improve resource utilization, and simplify the management of distributed applications, solidifying their role as an industry standard in both enterprise and open-source ecosystems. Furthermore, containers have also been applied in other contexts.

However, there are also new insights that are possible topics for the future projects. A particularly interesting finding of this project was the gap in research and practice with respect to security related themes in container-based development. During the project, candidate solutions were piloted by the companies, and the research institute produced a comprehensive literature review to probe the limits of present knowledge. However, it is clear that a dedicated research initiative in containers and security is needed.

Another finding that was evident was the considerable memory footprint of the containers. Developing techniques that enable using container-like technologies in devices that are restricted in memory remains a topic of future research. Some initial ideas have been proposed earlier[25], but they are at the level of prototypes at the moment, without industry-scale applications to test the approaches.

---

[25] Mäkitalo, N., Mikkonen, T., Pautasso, C., Bankowski, V., Daubaris, P., Mikkola, R., & Beletski, O. (2021, May). WebAssembly modules as lightweight containers for liquid IoT applications. In *International Conference on Web Engineering* (pp. 328-336). Cham: Springer International Publishing.

In addition to perfective research directions discussed above, a totally new direction that was introduced to the project during its execution is the use of generative AI in the context of containers. While techniques such as large language models were experimented in the project, the results are anecdotal at best, and creating something more comprehensive requires new research projects. At present, there is an ongoing ITEA project proposal in the process where such topics can be addressed. Furthermore, the doctoral pilot programme, initiated Summer 2024 has several students who are actively researching this topic.

## 6. Conclusions

In conclusion, project QLeap has produced new results with respect to using containers in software intensive business. To a degree, concerns raised in academic literature have been confirmed[26], but at the same time using containers has become a standard practice, especially in the context of cloud native systems[27]. The project also introduced some new directions for future work, the most important concerns being applying containers in systems that fall beyond the cloud native environment, such as small memory devices or regulated development. In addition, security concerns are still an important issue that has not been investigated thoroughly. Finally, techniques that are associated with generative AI that can be applied in container context are an important topic for future work.

---

[26] Mikkonen, T., Pautasso, C., Systä, K., & Taivalsaari, A. (2022, August). Cargo-cult containerization: A critical view of containers in modern software development. In 2022 IEEE International Conference on Service-Oriented System Engineering (SOSE) (pp. 93-98). IEEE.

[27] Oyeniran, O. C., Modupe, O. T., Otitoola, A. A., Abiona, O. O., Adewusi, A. O., & Oladapo, O. J. (2024). A comprehensive review of leveraging cloud-native technologies for scalability and resilience in software development. International Journal of Science and Research Archive, 11(2), 330-337.